# Causal Diagrams for Structural Engineers

M.Z. Naser, PhD, PE[1,2]
[1]School of Civil & Environmental Engineering and Earth Sciences (SCEEES), Clemson University, USA
[2]Artificial Intelligence Research Institute for Science and Engineering (AIRISE), Clemson University, USA
E-mail: mznaser@clemson.edu, Website: www.mznaser.com

## Abstract

Causal diagrams are logic and graphical tools that depict assumptions about presumed causal relations. Such diagrams have proven effective in tackling a variety of problems in social sciences and epidemiology research yet remain foreign to civil engineers. Unlike the traditional means of examining relationships via multivariable regression, causal diagrams can identify the presence of confounders, colliders, and mediators. Thus, this paper hopes to introduce the big ideas behind causal diagrams (specifically, directed acyclic graphs (DAGs)) and how to create and apply such diagrams to several civil engineering problems. Findings from the presented case studies indicate that civil engineers can successfully use causal diagrams to improve their understanding of complex causation relations, thereby accelerating research and practical efforts.

*Keywords*: Causality; Causal diagrams; Machine learning; Structural engineering.

## 1.0 Introduction

Causality aims to identify causal links by tying causes to effects where a cause is presumed to take place and trigger the occurrence of an effect [1]. More often than not, an effect is generated due to a number of causes. These causes could be independent or dependent on one another. Similarly, a cause could occur before other causes, and some may co-occur with other causes. One or more causes could have a common cause (confounder) or could, in fact, be the case of one another. A challenge then becomes to recognize a suitable means to tie causes to effect in a proper manner that preserves the reality of the phenomenon on hand.

Traditionally, engineers, and when first principles are available, can derive meaningful expressions. For example, the moment capacity of a W-shaped steel beam comprises the multiplication of the plastic modulus and yield strength of structural steel. When such principles are unavailable or complex, the engineers revert to statistical approaches – namely linear and multilinear regression. For example, the formula for evaluating the fire resistance of RC columns as adopted in the AS3600 code was also attained via multilinear regression [2]. Other formulae in structural engineering are also built on linear regression, such as those adopted in ASCE 29 [3].

In some instances, linear-like regression may not be able to realize proper formulae, and hence, the use of nonlinear regression analysis can be adopted. A common example of purely-based nonlinear regression is the formula created by Orangun et al. [4] to calculate bar development length in steel reinforcement embedded in concrete. Another example is the shear strength model for reinforced concrete beams as endorsed by the ACI committee 446 (Fracture Mechanics of Concrete) [5], as well as for FRP-reinforced concrete beams [6]. The use of nonlinear regression is quite common in seismic applications of structural design [7,8], as well as in examining size effects [9].

Bazant and Yu [10] note that while problems within the same domain may require a more rigorous statistical investigation (i.e., size effect) [10], some expressions (such as that for estimating the elastic modulus of concrete in terms of the compressive strength, etc.) can be arrived at empirically. Other expressions for member-level predictions were also arrived at empirically, such as that often used to evaluate the fire resistance of concrete-filled hollow steel columns [11] and in timber design [12].

The recent rise of nonparametric regression (such as symbolic regression) has also proved to be of merit in deriving engineering formulae. Such formulae maintain the transparent nature of displaying the dependent variables, as well as their relationships. However, this regression does not need to conform to the same assumptions used in traditional regression (linearity, uniformity, homoscedasticity, etc.). While symbolic regression can also yield formulae, yet the arrangement of the dependent variables may not mirror that derived via traditional analysis. Surprisingly, these models can also predict civil engineering phenomena with ease and accuracy that exceed those of their traditional counterparts [13].

A look into the above three variants of regression signifies that such variants are easy to use, well accepted, and most of all, interpretable. This interpretability arises on two fronts; 1) the fact that regression yields a formula that shows the relation between dependent variables, and 2) quantifies the contribution of each dependent variable upon the response of interest. Attaining a formula allows engineers to simply substitute this formula without the hassle of software or complex procedure.

Regression also extends to machine learning (ML) which can also be used to arrive at nonparametric models [14,15]. Yet, such models may not be as transparent (i.e., do not convert into a formula) [16,17]. The interpretation of such models is tedious and often incomprehensible unless supplemented with explainability measures. These measures can articulate the importance of model input, the relationship between each input due to arriving at correct and accurate predictions, etc. [18,19].

From an engineering sense, regression is often implicitly used to state that the dependent variables 1) predict or cause the phenomenon, and 2) this prediction and causation are in the form of the formula on hand. In reality, this may, and is likely, not be precise. For example, while linear regression can successfully create formulae to predict phenomena, the same formulas and models cannot be declared simply as causal.

This misconception could arise from pre-specifying the arrangement and form of the relationships adopted in such a formula. For example, the shear strength contribution of a typical concrete beam is given by the following formula:

$$V_c = 0.66\rho^{1/3}\sqrt{f'_c}b_w d$$

Where $\rho$, $b_w$ and $d$ represent the ratio of flexural reinforcement, the width of the web, and the depth of the beam, respectively.

The origin of formula comes from a comprehensive statistical analysis aimed at creating a formula to predict the shear strength of concrete [20]. The logic behind this formula matches that of our domain knowledge. For example, an increase in any of the above-listed dependent variables increases the shear strength.

Unlike regression and machine learning, graphs can also be used to describe relationships between dependent and independent variables. In such an exercise, the graphs can represent the relationships in a parametric or nonparametric manner as opposed to linear/nonlinear relations. More interestingly, given the graphic nature of graphs, these can also describe clear causal relations as well.

One of the earliest mentions of and use of graphs to display relationships is attributed to Sewall Wright [21], who established, mathematically, that a variable that resides in a graph causes another variable and not the other way around. Those graphs were called the "path diagrams" – a key part of the path analysis method. Using such graphs, Wright could articulate and defend causal assumptions mathematically and based on scientific grounds. Path analysis grows into factor analysis, and then structural equation modeling and much more in-depth reviews on the history and development of these methods can be found elsewhere [22–24].

## 2.0 An overview of regression

In regression, an engineer compiles a set of parameters identified empirically or from physics/domain knowledge to be tied to a phenomenon (i.e., temperature rise, deformation, stress, etc.). The hope is to create a tool (or a model or a formula) to predict or estimate the response from the compiled parameters. The creation of a tool must satisfy the requirement of the method used to create the tool.

For example, if linear regression is used to create a model, then this model must conform to the assumptions used in developing linear models. One of the first assumptions an engineer must follow is to set the predictors ($X_1$, $X_2$, $X_3$,.., $X_n$) and response ($Y$). Given the linear nature of the linear regression, these predictors can only be linearly tied to the response. In this particular example, the regression is multilinear (see Eq. 1). Frequently, in multilinear regression analysis, each relation of these independent variables is weighted solely to ensure the dependency of the predicted variable.

$$Y = \beta_0 + \beta_1 X_1 + \beta_2 X_2 + \cdots + \beta_n X_n + \epsilon \qquad \text{Eq. 1}$$

where $\beta_0$; is the y-intercept, $\beta_n$; is the regression coefficient of each independent variable $X_n$, and $\epsilon$ is the error.

There are more mechanical assumptions that an engineer must also satisfy when conducting linear and multilinear regression. These include:
- The presence of a linear relationship between the predictors and response.
- Any linear combination of the predictors must have a normal distribution.
- There exist no, or very little, correlation between the predictors.
- The independent of observations are statistically independent.
- Satisfying homoscedasticity (equal or similar variances in error among different groups).

Unfortunately, many of the above assumptions are left unchecked or implicitly assumed to be satisfied (please refer to Bazant et al. [25] for their analysis of data distribution in the shear database by ACI Committee 445 [25]) – as if they are not, then the application of traditional analysis is unlikely to be applied and engineers may not realize a working model.

By doing so, the engineer assigns that the predictors can, in fact, be thought of to predict, estimate or cause the response, with little regard to how true, such an assignment is, or to the fact that the predictors are independent of each other – and most importantly to the sufficiency the predictors provide (i.e., there are no other predictors that exist).

The above discussion on linear regression can be extended to nonlinear and symbolic regression (with suitable adjustments for the mechanical assumptions used in each method). The same can be carried over to machine learning, which has the least number of mechanical assumptions to satisfy.

### 3.0 Descriptive, predictive, and causal engineering queries

In general, queries pertaining to structural engineering problems can be roughly categorized into two classes, namely, 1) descriptive and 2) predictive. In hindsight, a third class also exists, referred to as causal, yet is rarely articulated as one. All of these are discussed herein.

The first type of query answer and describe engineering observations and their statistical relations. Engineers use such queries to establish a foundation for creating and tuning practical solutions (including formulae, methods, and theories). For example, this type of query can describe observations (i.e., on average, structures in coastal areas experience more corrosion issues than that inland) or compare the outcomes of experiments (e.g., all things being equal, larger columns have higher fire resistance ratings than smaller columns). The descriptive nature of these queries implies that descriptive questions cannot support counterfactuals (viz., they do not ask how a response would differ if some predictors were different).

On the other hand, predictive queries use data and observations to predict the response of interest given a set of *predefined* predictors. While descriptive queries are capable of addressing past and present observations, the second type of query can forecast future observations (whether at the individual or population level). The fact that these queries are built on predefined predicates implies that they, just like the first type of queries, cannot be used to identify or answer counterfactuals nor state the mechanisms for how and why the response occurs.

Finally, causal queries ask how a change(s) in structural response result from altering predictors. For example, a causal question can be: how would the midspan deflection curve changes if a beam depth is increased by 20%? Another question could be: all things being equal, does increasing a column's section or improving the material's strength leads to a much improved fire resistance? Answering the above questions can simply be done via conducting an experiment on a new deeper beam and by testing two identical columns (with the expectation of one being with a larger cross section and the other being fabricated with strong grade materials). Such tests can be costly yet doable. At the very least, cost-effective numerical models could be developed to answer such questions. However, such models would need to be validated against some ground truth, and if we lack a physical test, then the validity of such models may not truly, be established.

In some instances, if causal questions were raised at a system or community level, then physical testing nor numerical models can be used to answer such questions. Herein is where causality can come in handy.

**4.0 Causal diagrams**

As can be seen, one of the key drawbacks of statistical models is that they exemplify parametric assumptions that are not known to be correct and may well be incorrect. Another drawback of common statistical models is their incapability of capturing all types of assumptions (e.g., size and scaling effects, fire exposures other than standard fire (ASTM E119), etc.).

To transcend beyond mere correlations and associations and realize causal relations, an engineer must adopt causal assumptions. Such assumptions can be described via causal diagrams. These diagrams are graphical constructs that display the relationship between variables and responses in a web-like manner. This paper will focus on one type of causal diagram referred to as a Directed Acyclic Graph (DAG).

A DAG consists of nodes (variables) and arrows (directed edges) that tie the variables. The directed edges are *directed* because they follow time to order, and they are *acyclic* because they do not create any loops. DAGs visually state and encode causal relations and assumptions between the variables themselves and the response and hence allow for causal or counterfactual interpretation. A sequence of any unbroken route along or against arrows from variable to variable is called a path (see Fig. 1). A causal path is one that starts with the directed route of arrows leading tail to head from one variable to another (see A → C, also B→D→C).

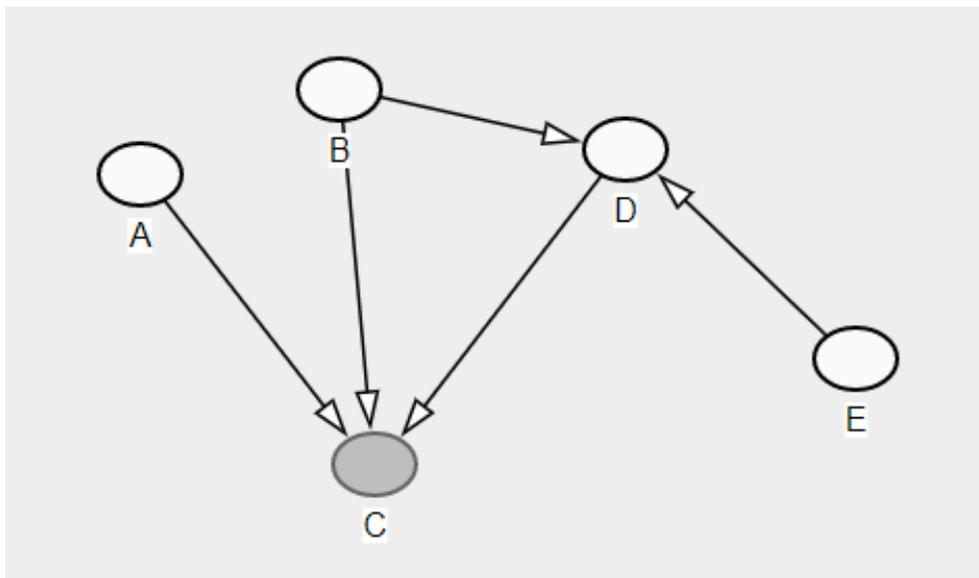

Fig. 1 A sample of a DAG

An intermediate variable along such a path that is not considered the change of interest nor the response is called a *mediator*. Three other types of intermediate variables are of interest, namely *confounders*, *colliders*, and *moderators* (see Table 1). An *instrument* is an external variable (not intermediate) – such as E. As one can see, any predictor in a regression analysis could be any of the above variables. In some instances, a variable can be one type of intermediate variable in one analysis and another in a different analysis – like D. With special treatment in each case, a traditional statistical and engineering analysis is likely to be faulty as it will be prone to bias and misinterpretation.

Table 1 Key definitions

| Term | Definition |
|------|------------|
| Confounder | A variable that has a causal effect on both the variables and/or variables and response, yet it is not affected by them (such as B in Fig. 1). Determining whether a variable is a confounder or not and whether if it should be controlled requires using logic (i.e., first principles, domain knowledge, etc.) as statistical analysis cannot determine this. |
| Collider | A variable that is a common effect of two other variables. The poor identification of a collider can generate distorted associations (such as D in terms of B and E in Fig. 1). |
| Mediator | A variable that causes another which in turn affects the response (such as B in Fig. 1). |
| Moderator | A variable that affects the strength and direction of a causal relationship. |
| Instrument | A variable associated with the exposure but is not related to the response except through its relationship with a variable. |

These are some examples of causal diagrams as applied to the following hypothetical structural engineering problems. Please note that none of the presented examples have a working mathematical/statistical/physical model, nor can they be solved by collecting and analyzing data alone. The discussion will be confined to 3- and 4-variable problems as a more in-depth discussion on higher order causal diagrams will be conducted in a future study. Interested readers are invited to review the following reviews on causal diagrams [26,27].

A case for a confounder in flood-prone zone due to strict building codes

Say that a dense zoning area ($Z$), such as a metropolitan in a flood-prone region, witnesses a limited number of structural failures ($F$) despite having a large number of structures ($N$) as compared to other zones within this region. A complimentary DAG for this case would be that shown in Fig. 2a. An explanation for this phenomenon would be that the zoning area of interest, unlike other zones, enforces strict codal provisions that enforce additional detailing and requirement to mitigate damage. As mentioned earlier, confounders are best identified via logic rather than statistical analysis – which would return empty if applied to the above.

A case for a collider in the rise of structural issues

Say that structural issues ($S$), like cracking, continue to rise in a given bridge population in one state. Such issues started to grow in the aftermath of lack of funding ($L$) as well as poor maintenance practices ($P$) adopted in that state. This can be depicted in Fig. 2b. as one can see, the two problems of lack of funding and poor maintenance collied to amplify the structural issues in bridges.

A case for a mediator (and instrument) due to a seismic event

Say that an earthquake occurs. During this earthquake ($E$), a building is severely damaged ($D$). This can turn into a mediator analysis in which the earthquake generates seismic loads ($M$) that damage the structure (see Fig. 2c). If such a structure is located in a seismically active region, this such a region is called an instrument (since it only affects the occurrence of the earthquake).

<u>A case for a moderator in structural fire engineering</u>

In this instance, the magnitude of deformation ($Y$) in a given structure can be amplified at elevated temperatures ($T$) and under the same level of loading ($G$) as that seen at ambient temperatures. Such larger deformations may not be observed under ambient conditions (see Fig. 2d). In this example, the more intense the fire, the more intense the temperature and the more severe the fire-induced degradation in materials take place, thus leading to larger deformations.

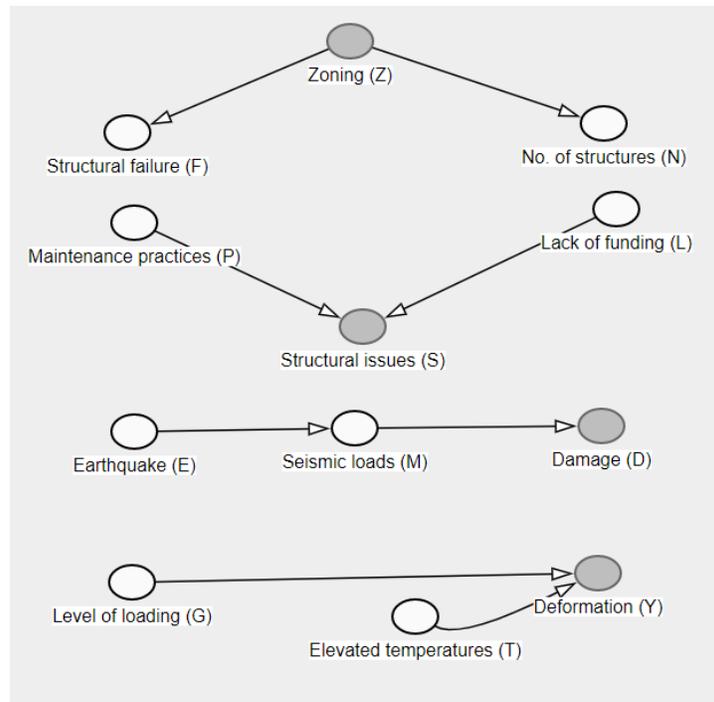

Fig. 2 Causal diagrams [Note: Top to bottom: a-d. Also, note that the slight curve in d is to distinguish the mediator effect.]

## 5.0 Conclusions

This paper presents a case for causal diagrams. Such diagrams can be of aid – especially in problems that lack physical representation or domain knowledge. A key motivation of this paper is to present such diagrams and how they can be implemented in structural engineering case studies. To go beyond regression, and improve our predictive capacity, adopting causal principles and diagrams can be of merit.

## Data Availability

Some or all data, models, or code that support the findings of this study are available from the corresponding author upon reasonable request.

## Conflict of Interest

The author declares no conflict of interest.